\documentclass[conference]{IEEEtran}
\IEEEoverridecommandlockouts
\usepackage{cite}
\usepackage{amsmath,amssymb,amsfonts}
\usepackage{algorithmic}
\usepackage{graphicx}
\usepackage{textcomp}
\usepackage{xcolor}
\usepackage{amsthm}
\def\BibTeX{{\rm B\kern-.05em{\sc i\kern-.025em b}\kern-.08em
    T\kern-.1667em\lower.7ex\hbox{E}\kern-.125emX}}

\newtheorem{definition}{Definition}

\newtheorem{challenge_base}{Challenge}

\newenvironment{challenge}{\begin{challenge_base}\upshape}{\end{challenge_base}}

\usepackage{subcaption}

\usepackage{url}

\begin{document}
\title{Low-Latency Stateful Stream Processing\\through Timely and Accurate Prefetching}

\author{
\IEEEauthorblockN{Eleni Zapridou}
\IEEEauthorblockA{
\textit{EPFL}\\
Switzerland \\
eleni.zapridou@epfl.ch}
\and
\IEEEauthorblockN{Anastasia Ailamaki}
\IEEEauthorblockA{
\textit{EPFL}\\
Switzerland  \\
anastasia.ailamaki@epfl.ch}
}

\maketitle

\begin{abstract}

Mission-critical applications often run ``forever'' and process large data volumes in real time while demanding low latency. To handle the large state of these applications, modern streaming engines rely on key-value stores and store state on local storage or remotely, but accessing such state inflates latency. As today’s engines tightly couple the data path with state I/O, a tuple triggers state access only when it reaches a stateful operator, placing I/O on the critical path and stalling the CPU. However, the keys used to access the state are frequently known earlier in the query plan. Building on this insight, we propose Keyed Prefetching, which decouples the data path from state access by extracting future access keys at upstream operators and proactively staging the corresponding state in memory before tuples arrive. This overlaps I/O with ongoing computation and hides the latency of large-state accesses. We pair Keyed Prefetching with Timestamp-Aware Caching, a cache-eviction policy that jointly manages previously accessed and prefetched entries to use memory efficiently. Together, these techniques reduce latency for long-running, real-time queries without sacrificing throughput.

\end{abstract}

\begin{IEEEkeywords}
Stream processing, State management, Low latency, Real-time
\end{IEEEkeywords}

\section{Introduction}

Mission-critical streaming applications such as fraud detection, payment systems, stock trading, and social networks operate under strict latency deadlines: as the time to react to a newly generated event increases, the usefulness of the output diminishes~\cite{kalavri-2020, tail-latency-socc, shadowsync}. Consequently, percentile latency governs performance, and production SLOs can be as low as 250 ms or even 50 ms~\cite{confluent-tail-latency, railgun}.

\begin{figure}[t]
 \centering
    \includegraphics[width=1\linewidth]{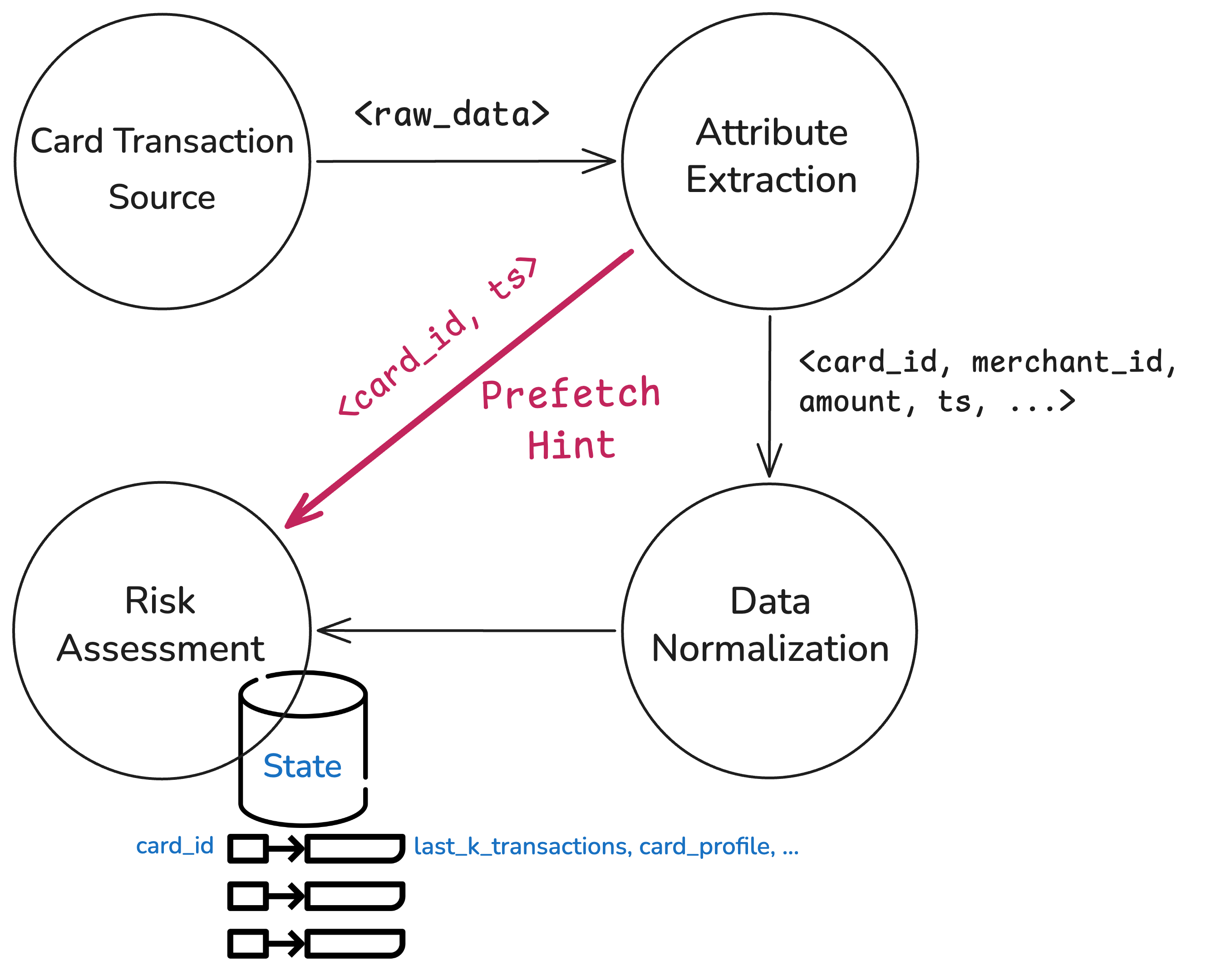}
    \caption{\textbf{Keyed Prefetching in fraud detection.} A card transaction goes through a pipeline of attribute extraction, data normalization, and risk assessment. Existing systems would start fetching the card-specific state (which grows with the number of cards) only when the tuple reaches the risk-assessment operator. Keyed Prefetching eliminates the state-access latency by accurately prefetching state in response to hints from the upstream attribute-extraction operator.}
	\label{fig:intro}
\end{figure}

Any non-trivial streaming computation relies on stateful operations, such as rolling aggregations, window contents, and timers. 
The growing data volumes and the shift to cloud-native deployments have led recent studies~\cite{benchmark-stream-stores, rhino, spe-evolution-survey, kalavri-2020, state-management-survey} and industrial products~\cite{Flink-v2-release, klaviyo, cloudera-tune-rocksdb, confluent-tune-rocksdb} to focus on managing state that resides outside the memory. For example, a major feature of Apache Flink 2.0 is a disaggregated state backend~\cite{Flink-v2-release}, while recent academic papers highlight the rise of large state and the need for efficient state access~\cite{flowKV2023, kalavri-2020}. 
However, accessing on-disk or disaggregated state significantly inflates tail latency~\cite{flowKV2023}.

As today’s engines tightly couple the data path with state I/O, a tuple triggers state accesses only when it reaches a stateful operator, placing I/O on the critical path and stalling the CPU. Approaches such as caching, asynchronous I/O, and software prefetching have traditionally been used to mitigate the cost of expensive state accesses. However, these methods alone fall short in addressing tail latency for streaming applications. First, caching policies reduce latency for hot keys, but tail latency remains high. 
Second, while asynchronous I/O can improve throughput by ensuring the CPU isn't idle during I/O, it does not eliminate the latency of fetching cold state; a tuple requesting cold state must still block until the I/O completes.
Finally, software prefetching aims to remove state access from the critical path, but existing prefetchers rely on predictable access patterns and are therefore a poor fit for streaming workloads.

Modern streaming engines manage state that exceeds memory capacity using persistent key--value stores~\cite{flink-state-management, structured-streaming-api, kafka-config, samza-linkedin}. Key--value stores are a natural fit, as systems typically organize state as key--value pairs to match their parallelism model, which relies on data partitioning~\cite{spe-evolution-survey}.

Although we cannot predict the key of a future tuple that will arrive at the streaming engine, within the engine, the keys used for state access by downstream operators are frequently known earlier in the query plan. 
Figure~\ref{fig:intro} illustrates a query from a fraud-detection system. The query receives card transactions as input, pre-processes them, and calculates the transaction's risk score by evaluating historical data. The state is organized by card IDs and grows as their number increases. This application has tight latency requirements and must respond quickly to all incoming transactions, even if they concern a card ID that has not been seen in a while. In existing systems, the operator calculating the risk score would read its next input tuple and then fetch the necessary state, inflating latency. However, the card ID attribute used to access the state is already known in the attribute-extraction operator.

This paper introduces \textbf{Keyed Prefetching}, a technique that decouples state I/O from the data path, treating tail latency as a first-class objective. Keyed Prefetching uses \emph{lookahead} operators as a mechanism for informing downstream stateful operators of the keys they will soon need to access. These stateful operators can then fetch the state as soon as a key is known in the query plan---before the tuple reaches the operator. \emph{Lookahead} operators emit accurate prefetch hints, the state is fetched asynchronously, and tuples find their state hot, thereby avoiding CPU stalls and reducing end-to-end latency. 
To jointly manage prefetched and previously accessed state, we propose \textbf{Timestamp-Aware Caching}, which relies on event timestamps: the timestamp of a prefetch hint encodes when a key is expected to be accessed, while the timestamp of an already accessed entry records its most recent access time. This way, timestamps provide a single ordering signal that naturally ranks both entry types.
To minimize tail latency while keeping overheads low, Keyed Prefetching (i) omits hints for very frequent keys that are unlikely to lead to cache misses, (ii) adjusts the selected \emph{lookahead} at runtime to ensure hints are accurate and timely, and (iii) ensures writes are not on the critical path of execution and can be performed asynchronously.
Overall, Keyed Prefetching accurately prefetches state, uses memory efficiently, and enables streaming applications to scale to larger states without sacrificing timely responses. In addition, its design is general and can be used with any underlying key--value store that stores data locally or remotely.

This paper makes the following contributions:

\begin{itemize}

 \item We minimize I/O delays on the critical path of execution by decoupling the data path from state accesses. Our experiments show that accessing state as soon as the key is known hides I/O costs, thereby minimizing tail latency.

 \item We propose Keyed Prefetching, a mechanism that exploits query-plan semantics to prefetch state based on information from upstream operators, hiding state access.
 
 \item We propose Timestamp-Aware Caching, which employs event timestamps to seamlessly operate over both prefetched and previously accessed state. By balancing recent and anticipated future use, this policy uses memory efficiently for both state types. In our experiments, using  Keyed Prefetching with Timestamp-Aware Caching reduces p999 latency by $1.34-11\times$ without sacrificing throughput.

 \,

\end{itemize}

\section{The Tension Between State Volume and Tail Latency}
\label{sec:background}

\begin{figure}[t]
 \centering
    \includegraphics[width=1\linewidth]{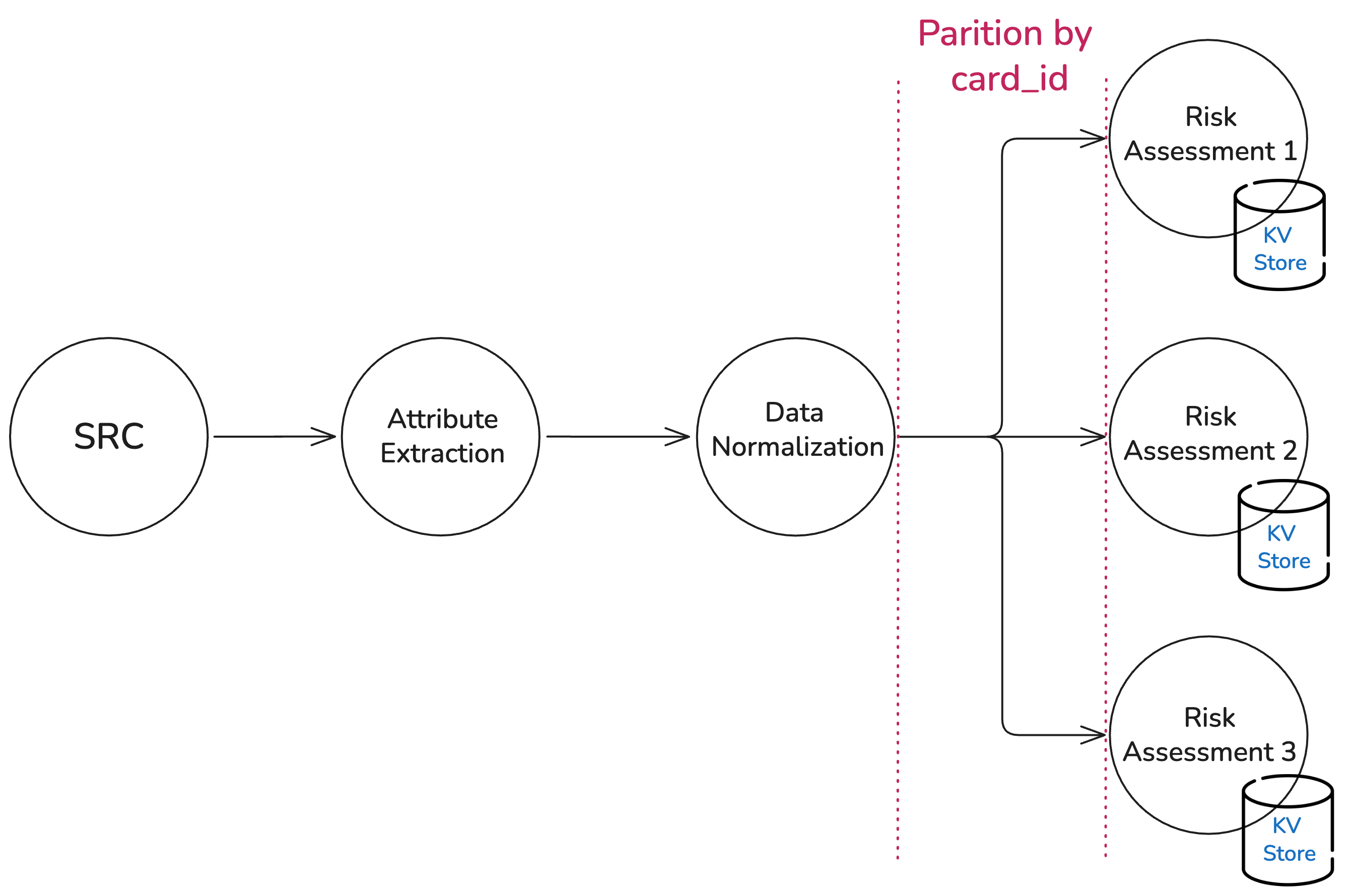}
    \caption{A parallel dataflow graph corresponding to the query of Figure~\ref{fig:intro} with embedded key--value stores for subtasks of stateful operators. Tuples between the data-normalization and risk-assessment operators are partitioned by the card ID attribute, and the same attribute is used to partition the state of the risk-assessment operator. For simplicity, we present the source, attribute-extraction, and data-normalization operator with a single subtask.}
	\label{fig:stateful-parallel-dag}
\end{figure}

The stream processing model assumes a dataflow, typically represented as a directed acyclic graph (DAG), where nodes represent operators and edges represent data streams. We adopt the tuple-at-a-time processing model as it is more suitable for low-latency applications~\cite{discretized-zaharia}. Our method can be extended to the micro-batch processing model by extracting and forwarding prefetch hints for an entire batch at once.
Using the tuple-at-a-time processing model, each operator has an input and an output queue and operates at the tuple granularity: it pulls a tuple from the input queue, processes it individually, and enqueues it to the output queue. 

\textbf{Streams \& Windows}. A stream $S$ comprises an infinite sequence of records that obey a partial order. We also assume that the order $t$ of a tuple $e$ in the stream is explicitly expressed as an attribute of the tuple, i.e., $e_t=(t,\bullet)$.

\textbf{Pipelined operators}. When data does not need to be shuffled between two operators, systems typically \textit{pipeline} them. Pipelined operators are therefore a single node in the graph and, from the system's perspective, are considered as one operator. 

\textbf{Parallel dataflows}. In distributed stream processing systems, each operator has multiple deployed instances, which we call \emph{subtasks}.
For example, in Figure~\ref{fig:stateful-parallel-dag}, there are three subtasks for the risk-assessment operator.
Each record is typically sent to the next subtask compounded with other records in a network buffer, the smallest communication unit for communication between subtasks. Smaller buffer sizes and tighter buffer timeouts trade off throughput in terms of records per second for lower latency~\cite{low-latency-config-flink}.
To employ parallelism and split tuples between subtasks, systems rely on data partitioning. Tuples are routed based on a set of specific attributes that act as partitioning keys. In the example of
Figure~\ref{fig:stateful-parallel-dag}, tuples that have the same card ID should be routed to the same subtask. To distinguish partitioning keys from the rest of the attributes, we denote tuples as $e_t=(t,k,v)$, where $k$ is the key.

\textbf{Stateful dataflows.}
Non-trivial streaming queries maintain state for their operations (e.g., windowed aggregation, join, deduplication) and window timers/triggers. The state is partitioned using the same key used to partition the data. In the example of Figure~\ref{fig:stateful-parallel-dag}, the state of the risk-assessment operator is partitioned by card ID, and each subtask has its own state.
This design choice promotes scalability, state locality, and isolation between subtasks.
As data volume and query complexity increase, the state can exceed memory capacity and spill to local storage or, in modern containerized cloud deployments~\cite{Flink-v2-release}, to disaggregated storage. In real-world applications, state can reach hundreds of terabytes~\cite{klaviyo, Flink-v2-release}. 
To manage state at this scale, modern streaming systems rely on persistent key--value stores. The most prominent store is RocksDB\cite{rocksdb}, used by open-source systems such as Apache Spark Structured Streaming\cite{structured-streaming-api}, Apache Flink~\cite{flink-state-management}, Apache Kafka\cite{kafka-config}, and Apache Samza~\cite{samza-linkedin}.

\subsection{Large states hurt tail latency}

The end-to-end latency for an output record is shaped by operator processing times, queuing at operator boundaries, and inter-operator communication. As data flows through the DAG, an operator consumes an input tuple and produces one or more transformed tuples---or drops it entirely. Latency then accumulates along the specific path taken by these produced tuples.

Let $G = (V, E)$ be the dataflow graph. For a particular output record $e_t$, let the executed operator path that produced $e_t$ be $\pi(e_t) = (v_1, v_2, ..., v_K)$. For each stage $i$, $q_i(e_t)$ is the time spent queuing before operator $v_i$ (i.e. between $v_i$ and $v_{i-1}$ or between $v_i$ and the data source if $i=0$), $p_i(e_t)$ is the in-operator time, i.e., the time spent within the operator for processing and potential I/O and $c_i(e_t)$ the communication time for the tuple to travel to the next operator. Then, the end-to-end latency of $e_t$ is:
\begin{equation}
    L(e_t) = \sum_{i=1}^K(q_i(e_t) + p_i(e_t)) + \sum_{i=1}^{K-1}c_i(e_t)
\end{equation}

Some input tuples might not result in any output tuples; however, they can still affect the latency of later output tuples---e.g., by increasing their queuing time. 

If the processing of a tuple by a stateful operator requires cold state, the operator gets blocked waiting on I/O. This increases the latency of the particular tuple (by increasing $p_i$) but also that of future tuples, which must be queued longer (increasing $q_i$). \emph{Therefore, tail latency often reflects cold accesses from multiple preceding records and thus can far exceed the time to process a single cold-key record.}

Caching is a standard technique to mitigate the overhead of accessing large or disaggregated state. Streaming engines place in-memory caches in front of the embedded key–-value stores with LRU and Clock being typical choices for the eviction policy~\cite{flink-configuration, rocksdb-block-cache}. Caching reduces average latency for hot keys, but cold‑key accesses still inflate the tail percentiles. Moreover, studies in streaming workloads show that keys are highly ephemeral and have a short time-to-live, limiting the effectiveness of purely reactive caches~\cite{benchmark-stream-stores}.

Asynchronous I/O can increase throughput by overlapping processing with state access. However, a tuple whose processing requires cold state will still need to wait for the state to be fetched. Consequently, for low percentile latency, asynchronous I/O alone does not suffice. 
Streaming engines provide limited support for asynchronous I/O~\cite{flink-asyncIO}.

\subsection{The challenges of prefetching streaming state}
\label{sec:challenges}

Prefetching aims to move state accesses earlier in time, before a state element is actually needed, and this way to overlap them with upstream computation. 
However, the dynamic and unpredictable nature of data streams amplifies the traditional prefetching challenges. We use the term ``prefetching'' in the broad sense of fetching state earlier than it will be consumed. In the context of the streaming dataflows this paper addresses, this translates to proactively fetching operator state from the persistent key--value store to the cache.

\begin{challenge}
\emph{\textbf{What to prefetch?}}
The keys a stateful operator will access depend on input data being generated in real time. Existing prefetchers~\cite{prefetching-joins, btreeprefetch, fractalPrefetch, 10.1145/237090.237190, 10.1145/143371.143488} rely on detecting workload patterns. However, streaming data follow evolving trends that can change at any time. Crucially, for streaming applications, such as fraud detection, alerts, and monitoring, moments that diverge from history and, therefore, when the workload is shifting, are particularly interesting and precisely when we cannot afford performance drops.
\end{challenge}

\begin{challenge}
\emph{\textbf{When to prefetch?}} If we prefetch too late, we fail to hide state access. On the contrary, prefetching too early introduces memory pressure and risks evicting prefetched elements before they are even used. Streaming workloads exacerbate this classical problem: state elements have short time-to-live, and spatial locality is limited as consecutive events may hit very different keys in state~\cite{benchmark-stream-stores}.
Therefore, to efficiently use the cache, we must update the elements cached quickly.
\end{challenge}

\begin{figure}[t]
 \centering
    \includegraphics[width=1\linewidth]{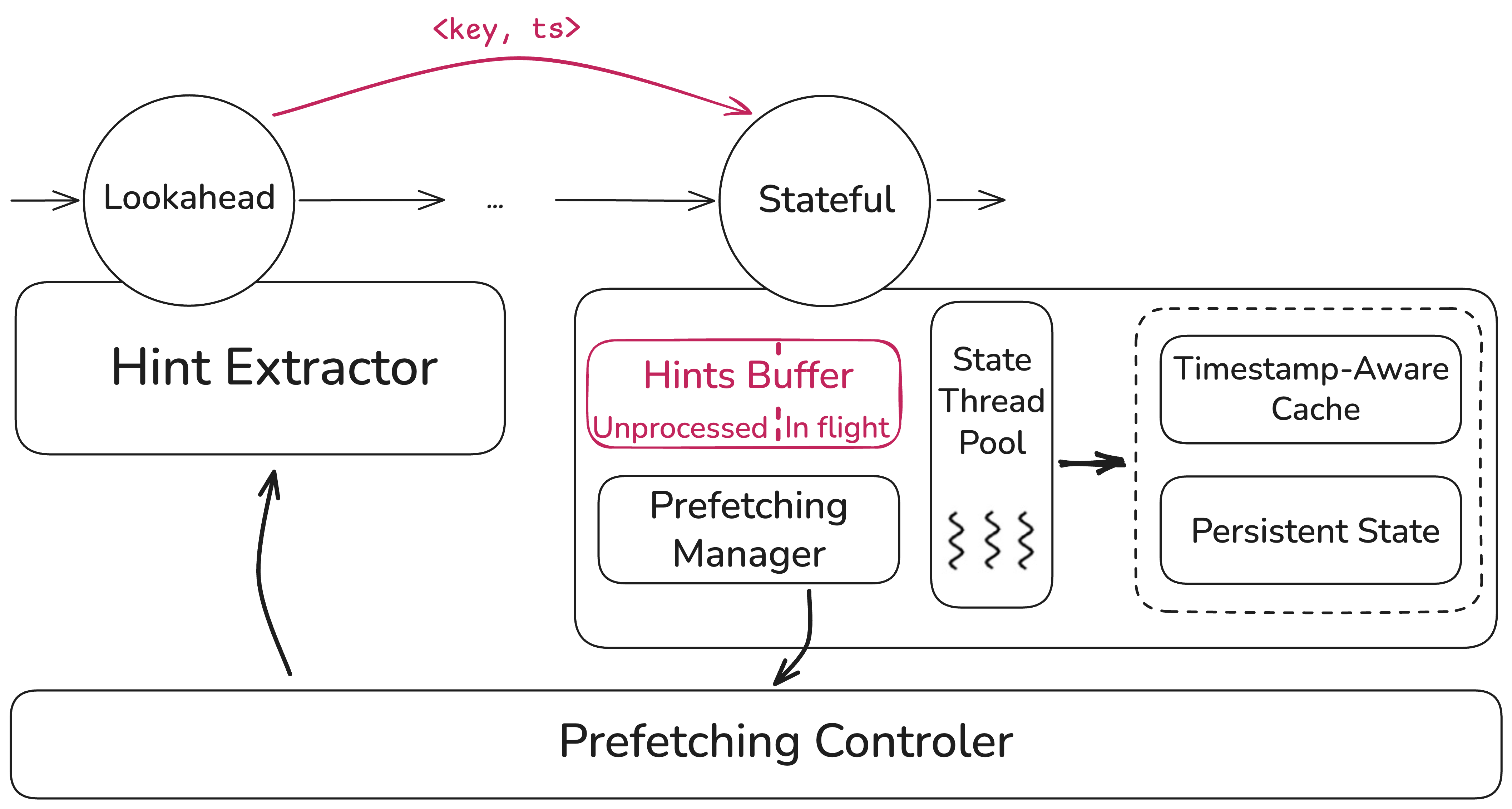}
    \caption{Keyed Prefetching Overview}
	\label{fig:overview}
\end{figure}

\begin{challenge}
\emph{\textbf{Efficient cache utilization.}} With prefetching, the engine needs to cache two classes of entries: (i) prefetched state and (ii) recently accessed state. Favoring one can starve the other: over‑admitting prefetched keys evicts frequently-accessed entries, while excessively limiting the number of keys that can be prefetched causes failure to remove I/O from the critical path.
\end{challenge}

\begin{challenge}
\emph{\textbf{Prefetching overheads vs low latency.}} 
Prefetching introduces overheads: (i) the computational cost of predicting future accesses, and (ii) the memory and I/O bandwidth wasted on mispredictions.
Under strict tail-latency constraints, minimizing these overheads is critical.
\end{challenge}

\section{Decoupling State Access from the Data Path}
\label{sec-overview}

 Streaming systems couple state access with the data path. An operator only accesses the state necessary to process a tuple after the tuple has been processed by all upstream operators. This coupled design leads to inflated tail latencies, as a single tuple might need to pass through multiple stateful operators and can also be delayed by state accesses from preceding tuples due to queuing.

The information that determines which state must be accessed does not need to follow the same path as the information necessary for processing a tuple and incorporating it in the output. 
Using this insight, we can decouple state access from the data path. This way: (i) the information relevant for state accesses travels faster and is used to fetch the corresponding state ahead of time, and (ii) the data path can also move faster as it does not get blocked on I/O. This decoupling enables accurate prefetching, which, instead of being speculative, relies on accurate information originating from earlier in the query plan.

Next, we present how, based on this insight, we design Keyed Prefetching, a mechanism that proactively fetches state and efficiently utilizes the cache to achieve low tail latency. Keyed Prefetching targets operators that need to access keyed state upon the arrival of an input tuple. This includes read-modify-write operators---such as associative and commutative aggregations and incremental processing, such as incremental joins---and read-only operators---for example, tuple enrichment.

\subsection{Overview of Keyed Prefetching}

Figure~\ref{fig:overview} illustrates the architecture of Keyed Prefetching and how it interacts with the processing layer and the persistent state. 
The \emph{Prefetching Controller} is a centralized component that stores the IDs of the candidate \emph{lookahead} operators for each stateful operator and, upon request from stateful operators, adjusts the selection of the \emph{lookahead}. 
The \emph{Hint Extractor} inside lookahead operators is responsible for extracting hints, which then get forwarded to the stateful operator. Stateful operators have a \emph{hints buffer} which temporarily stores hints before they are handled and the \emph{Prefetching Manager}, which is responsible for monitoring runtime statistics and requesting to switch to a different \emph{lookahead} if hints are not accurate or do not arrive early enough to hide state access. The stateful operator also includes the \emph{State Thread Pool}, which asynchronously performs I/O operations between the \emph{Timestamp-Aware Cache} and persistent state. 

The design of Keyed Prefetching addresses the challenges mentioned in Section~\ref{sec-overview}. First, upstream operators inform stateful ones exactly which keys will need to be accessed soon (Challenge 1). Second, by adjusting the selected \emph{lookahead} operator at runtime, Keyed Prefetching selects the one that provides the most timely hints (Challenge 2). Third, our \emph{Timestamp-Aware Cache} jointly handles both previously accessed and prefetched state and thus, the percentage each utilizes automatically adjusts to the workload, as described in Section~\ref{sec:timestamp-aware-cache} (Challenge 3). Finally, by using exact information from upstream operators, Keyed Prefetching eliminates prefetching mispredictions and naturally detects what should be fetched. Additionally, to further reduce the overhead, we minimize the hints that should be sent by an upstream operator (Section~\ref{sec:keyed-prefetching-details}) by omitting hints for very frequent keys, and we design the \emph{Timestamp-Aware Cache} to keep write operations outside of the critical path of execution.

\section{Prefetching Streaming State}
\label{sec-method}

Prefetching streaming state from the backend to memory requires accurate prefetching hints, fetching state in a timely manner to effectively hide I/O without polluting the cache with state that is needed much later in the future, and balancing the available memory between previously-seen and prefetched state while keeping overheads low even for percentile latencies. Keyed Prefetching prefetches state accurately and timely by relying on query-plan semantics and selecting the operators that sent prefetch hints at runtime. The design of Keyed Prefetching (Figure~\ref{fig:overview}) consists of four main components: (i) the \emph{Prefetching Controller} oversees the entire query and is responsible for adjusting the choice of operators that produce the \emph{prefetch hints} and enabling the prefetching mechanism, (ii) \emph{lookahead} operators use a \emph{Hint Extractor} to extract \emph{prefetch hints} and forward them to the stateful operator, (iii) the stateful operators have a \emph{Prefetching Manager} which handles \emph{prefetch hints} and decides when to trigger a fetch request and (iv) the \emph{Timestamp-Aware Cache} is placed in front of the persistent state and is responsible for caching both previously-accessed and prefetched state.

\begin{definition} [Lookahead operator] We call the operator sending prefetch hints to a steteful operator a \emph{lookahead} operator. 
\end{definition}

\begin{definition} [Prefetch hint]
    A prefetch hint is a tuple $h_t = (k, t)$ which travels between a lookahead and a stateful operator via a specialized channel and informs the stateful operator that the state of key $k$ will need to be accessed at event time $t$.
\end{definition}

In this section, we first present the function of the \emph{Prefetching Controller} (Section~\ref{sec:prefetching-controller}) and the \emph{Hints Extractor} (Section~\ref{sec:lookahead}), and then detail the operations of the \emph{Prefetching Manager} (Section~\ref{sec:keyed-prefetching-details}) and the \emph{Timestamp-Aware Cache} (Section~\ref{sec:timestamp-aware-cache}). 

\subsection{Prefetching Controller}
\label{sec:prefetching-controller}

\paragraph{Identifying candidate lookahead operators} 

Once a query is submitted, the \emph{Prefetching Controller} identifies for each stateful operator $v_S$, candidate operators that can provide prefetch hints and keeps them sorted based on their order in the query plan (from the one closest to the source to the one closest to the stateful operator). An operator is a candidate \emph{lookahead} for $v_S$ if the following conditions hold:
\begin{enumerate}
    \item it is an upstream operator to $v_S$
    \item it is able to extract from the payload of its output tuples the key that $v_S$ is using to access state
\end{enumerate}

Initially, no operator sends prefetching hints to $v_S$ as the state might be fitting in the cache. At runtime, if $v_S$ observes cache misses, it then requests the \emph{Prefetching Controller} to activate one of the candidate \emph{lookahead} operators so that it begins producing \emph{prefetch hints}. 

Different candidate \emph{lookahead} operators can provide \emph{prefetch hints} of different quality. For example, an operator that is placed early in the query plan, many stages away from $v_S$, would send hints to $v_S$ earlier and provide more time to prefetch state, while an operator closer to $v_S$ is more likely to observe the same key distribution as $v_S$ as operators between a candidate \emph{lookahead} and  $v_S$, such as filters and joins, can change the distribution of the state access key.

To achieve timely and accurate prefetching, our goal is to choose the \emph{lookahead} operator such that it is the last candidate in the query plan that:
\begin{enumerate}
    \item observes the same key distribution as $v_S$
    \item provides enough prefetching time
\end{enumerate}

We achieve this via a lightweight adaptive policy driven by runtime information. Initially, the \emph{Prefetching Controller} selects the first candidate operator as the \emph{lookahead}. At runtime, if the \emph{Prefetching Manager} of the stateful operator $v_S$ observes that the distribution between the stateful and \emph{lookahead} operators is different, it requests from the \emph{Prefetching Controller} to switch to a later \emph{lookahead} operator. The \emph{Prefetching Controller} then discards the current and all upstream \emph{lookahead} operators from the list of candidates before switching to a later one.
We completely discard \emph{lookahead} operators that see a different distribution of the state access key to prioritize hint accuracy. 
While our current implementation enforces this 0\% prefetch-miss threshold, relaxing it could benefit some workloads. It would introduce a trade-off between having a larger prefetch window and the overhead of prefetch misses. Allowing a small ratio of prefetch misses would enable the use of lookahead operators positioned earlier in the query plan, increasing the time available to hide I/O, which could be beneficial when the most accurate lookahead operator does not provide sufficient slack. However, this comes at a cost: fetching unused state consumes I/O bandwidth and risks polluting the cache with entries that are never accessed. We leave exploring this trade-off as future work.

The stateful operator can also request a different lookahead operator based on the prefetch slack---the time between a hint's arrival and the corresponding tuple's arrival. We detail this process in Section~\ref{sec:keyed-prefetching-details}.

\subsection{Lookahead operator \& Hints Extractor}
\label{sec:lookahead}

\paragraph{Communicating prefetching hints}
The Hints Extractor is a lightweight routine embedded in the selected \emph{lookahead}: for each output record $e_t=(t,[payload])$, it extracts from the payload the state-access key $k$ of $v_S$ and emits a \emph{prefetch hint} $h_t=(k, t)$ on a side channel toward $v_S$. This side channel is an additional edge in the dataflow query graph---thereby keeping the execution graph a DAG---so existing streaming engines can handle it natively. We route hints using the same partitioning function as the data stream, ensuring each $v_S$ subtask receives hints only for its local partition; this is possible because state and data in $v_S$ are partitioned the same way. The stateful operator thus receives two types of input: the regular data input and a hint input. Upon hint arrival, the Prefetching Manager buffers or acts on the hint (Section~\ref{sec:keyed-prefetching-details}), and the Timestamp-Aware Cache records anticipated future use (Section~\ref{sec:timestamp-aware-cache}). This design decouples state access from the data path while remaining non-invasive to the engine’s scheduler, backpressure, and checkpointing mechanisms. 

\paragraph{Minimizing hints}
The goal of Keyed Prefetching is to inform stateful operators of keys that are currently not in memory and will be accessed soon. Streaming workloads are typically highly skewed~\cite{Chi2018, skewedData}. 
Frequent keys will anyhow be present in the cache and will not incur expensive accesses. Therefore, to lower the communication overhead between the \emph{lookahead} and stateful operator and to reduce the cost of processing hints in the stateful operator, the \emph{lookahead} can omit hints that concern very frequent keys.

Specifically, the lookahead operator employs a filter that detects frequent keys and extracts hints only for keys that do not pass the filter.
 A coarse filter that detects very hot keys is sufficient; we do not need to rank all keys precisely. 

We use a tiny  Count–Min Sketch (CMS) with saturating counters. This design has been used before for cache admission~\cite{tinyLFU} and is very suitable for our use case because it is very lightweight while keeping statistics fresh, which is necessary for evolving streaming data. The sketch has $d$ rows and width $w$; each cell is a $b$-bit counter. On seeing key $k$, the \emph{Hints Extractor} updates the sketch:

\begin{equation*}
    \forall {i \epsilon \{1,\dots,d\}}: C[i,h_i(k)] \leftarrow min\{C[i,h_i(k)]+1, 2^b - 1\}
\end{equation*}

where $h_i$ are $d$ independent hash functions. Every $\Delta$ (either wall-clock time or a fixed number of records), we \emph{age} the structure by dividing each counter by 2 using integer division (i.e., right shift by one bit): $C[i,j]\leftarrow C[i,j] \gg 1$. This aging process keeps frequency estimates fresh and bounds the counters, keeping the memory footprint tiny.

A key is classified as \emph{hot}—and is omitted from hint emission—iff \emph{all} its $d$ touched counters exceed a threshold $T$:

\begin{equation*}
    \bigvee_{i=1}^d C [i,h_i(k)] \geq T
\end{equation*}

The filter quickly detects very hot keys, which are highly likely to be present in the stateful operator's cache, and so, they will not produce hints.
Each tuple costs $O(d)$ hash+increment operations; the state footprint is tiny for practical $(d,w,b)$, and the aging pass is inexpensive given the tiny array.

\subsection{Prefetching Manager}
\label{sec:keyed-prefetching-details}

\paragraph{Handling prefetch hints} 
The \emph{prefetching manager} runs in the stateful operator and maintains a two‑stage hints buffer. The buffer stores, for each state access key, the latest event‑time timestamp seen so far, and deduplicates keys so that there is at most one entry per key. The buffer is split into two sections: (i) unprocessed: keys for which a hint has been received but no fetch has been issued yet, and (ii) in‑flight: keys for which an asynchronous fetch has been initiated but has not completed yet.

\textbf{Arrival of hints.} As the \emph{lookahead} operator (Section~\ref{sec:lookahead}) does not send hints for very hot keys and routes hints by key partition, every subtask of the stateful operator only sees hints for its local partition and predominantly for cold keys.
When a hint $h_t = (k, t)$ arrives, the manager first consults the \emph{Timestamp-Aware Cache}. If the key is already cache-resident, the \emph{Prefetching Manager} just notifies the \emph{Timestamp-Aware Cache} of the anticipated future use at timestamp t so that the cache updates the entry's metadata (see Section~\ref{sec:timestamp-aware-cache}). If the key is not cached, the \emph{Prefetching Manager} checks if the key is already in the buffer. If that is the case, it updates the timestamp to $t \leftarrow max(t,t_{old})$, otherwise it inserts the key with the timestamp into the unprocessed section of the buffer.

\textbf{State Thread Pool.}
A small pool of asynchronous I/O workers repeatedly select keys from the unprocessed part of the buffer, move them to in‑flight, and fetch the corresponding state from the backend. Once the state is fetched, the hint is removed from the buffer, and its state is inserted in the cache.

\textbf{Arrival of regular data.} When a regular tuple arrives, if the state is already in the cache, the operator proceeds normally with processing it. In the unlikely event that the state is still cold (e.g., a hint arrived late or prefetching has not completed), the operator issues a fetch for the key (without duplicating any ongoing prefetch).

\paragraph{Selecting the lookahead operator}
Our adaptive policy for selecting the \emph{lookahead} operator combines two complementary runtime mechanisms to ensure both timely and accurate prefetching. First, the \emph{Prefetching Manager} based on runtime statistics selects from the current candidate operators which one offers the most timely hints---hints should arrive early enough to hide I/O but not too early to inflate buffering overheads. Second, the manager also keeps track of hint accuracy and identifies candidate \emph{lookahead} operators that do not provide accurate hints.

\textbf{Adapting prefetching timing.} To decide where \emph{prefetch hints} should originate, we use a low‑overhead marker mechanism that directly measures the prefetching slack each candidate \emph{lookahead} provides to the stateful operator $v_S$. As mentioned in Section~\ref{sec:prefetching-controller}, the goal is to pick the latest \emph{lookahead} that (i) leaves enough time to hide state I/O and (ii) observes the same key distribution as $v_S$.
Markers are a coordination mechanism in streaming engines---used, for instance, in reconfiguration, and snapshotting\cite{Chi2018, carbone2015lightweightasynchronoussnapshotsdistributed, fries}---to trigger control-plane actions without interfering with data processing. 
Keyed Prefetching uses markers to efficiently calibrate prefetch timing.

Every $T$ time units, the \emph{Prefetching Controller} injects a marker into all source operators. Markers then propagate through the dataflow, interleaved with normal data messages. 
When a candidate \emph{lookahead} operator receives a marker: (i) it forwards the marker downstream on its regular output channel, preserving order, and (ii) it sends the same marker with its operator id to the stateful operator $v_S$ through the hint side channel.
Operators that do not produce hints simply forward markers downstream.
When $v_S$ receives a maker through the hint channel, it records the arrival time $t_{hint}(L_i)$ associated with the originating candidate \emph{lookahead} $L_i$. When the same marker later arrives through the regular data channel, $v_S$ records the arrival time $t_{data}$. Once all markers for this round have arrived, $v_S$ computes for each candidate:

\begin{equation*}
    G_i = t_{data} - t_{hint}(L_i)
\end{equation*}
which represents the slack window the candidate $L_i$ provides (the time between a hint's and the corresponding tuple’s arrival at $v_S$).
Because markers traverse the same path as normal data and hints, $G_i$ captures end-to-end queuing and backpressure effects, providing a precise measure of prefetching slack.

The \emph{Prefetching Manager} maintains the percentile (e.g., p99) of the measured $G_i$ values and also tracks the state access percentile latency $F$.
Based on these statistics, the \emph{Prefetching Manager} selects the \emph{lookahead} operator that satisfies the following condition:
 \begin{equation*}
    L^* = max_i \{L_i | p_{99}(G_i) \geq p_{99}(F) + \gamma\}     
 \end{equation*}
where $\gamma$ is a small safety margin (e.g., a few milliseconds).

Intuitively, $L^*$ is the latest lookahead that still leaves enough time—at the 99th percentile—to complete prefetching before the corresponding tuple reaches $v_S$.

\textbf{Detecting distribution mismatch.}
A candidate \emph{lookahead} may observe a different distribution over the state access key than $v_S$, leading to the prefetching of irrelevant states.
To detect this, the \emph{Prefetching Manager} monitors the fraction of prefetching misses, i.e., state entries that were fetched but never accessed.
If this prefetch-miss ratio exceeds a threshold, the \emph{Prefetching Manager} requests from the \emph{Prefetching Controller} a later lookahead closer to $v_S$, whose emitted hints are more likely to better reflect the operator’s current access pattern. In this work, we set this threshold to $0\%$, and therefore Keyed Prefetching converges to using \emph{lookahead} operators that observe exactly the same distribution as the stateful operators.

\subsection{Timestamp-Aware Cache}
\label{sec:timestamp-aware-cache}

\begin{figure}[t]
 \centering
    \includegraphics[width=1\linewidth]{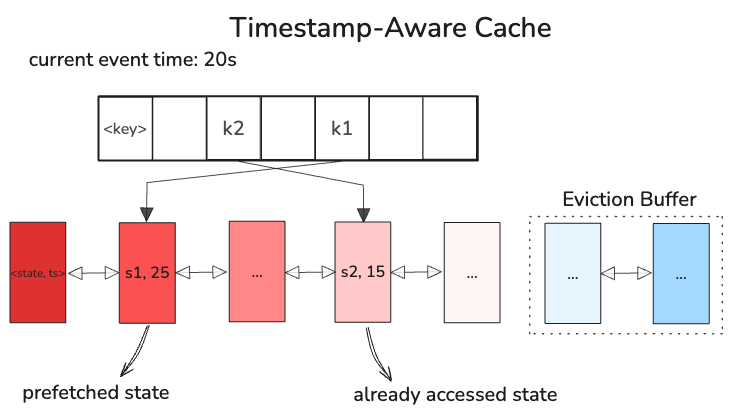}
    \caption{Timestamp-Aware Cache Design}
	\label{fig:tac}
\end{figure}

The \emph{Timestamp-Aware Cache} (TAC) manages both prefetched and previously accessed state uniformly. Timestamps—already present in streaming records—provide a natural ordering signal allowing the cache to reason about both state types jointly. TAC's design is illustrated in Figure~\ref{fig:tac}.

To manage memory efficiently, the TAC prioritizes evicting entries with the oldest timestamps. Each cache entry corresponds to a key $k$ and holds $(k, s_k, t_k)$ where $s_k$ is the state object and $t_k$ its timestamp. The cache assigns timestamps differently depending on the origin of the entry: \textbf{Previously-accessed state:} When $v_S$ reads or updates the state of key $k$, the cache sets $t_k$ to the \emph{time of last access} (event time). Among previously accessed states, the TAC thus behaves like a Least-Recently-Used (LRU) cache based on event timestamps. \textbf{Prefetched state:} When a state entry gets prefetched into the cache, its timestamp $t_k$ is set to the event-time of the hint that triggered the fetch, which is \emph{the timestamp of when the state is expected to be used}.
Because hints originate from tuples of upstream operators that will reach $v_S$ later, their timestamps are ahead of the operator’s current event time.  
This unified timestamp ordering naturally positions prefetched entries ``in the future'' relative to the currently active state. 
When a new hint for an already-cached key arrives, the cache updates the entry’s timestamp to the timestamp of the hint. This operation effectively renews the entry’s predicted relevance: it remains in the cache longer, preventing premature eviction when the key is expected to be accessed again soon.
Ultimately, evicting the smallest-timestamp entries naturally preserves recently accessed or soon-to-be-used state entries.

To support fast lookup, ordered eviction, and non-blocking writes, the TAC employs a three-data-structure design:
\textbf{Hash map: } A hash map provides $O(1)$ access and update by key. It stores pointers to the nodes of the ordered list and the \emph{eviction} buffer.
\textbf{Doubly-linked list ordered by timestamps: } All entries are sorted in descending order of $t_k$. This structure allows for fast evictions---simply removing the tail of the list, which has the oldest timestamp. \textbf{Eviction buffer:} When an entry is selected for eviction, the TAC checks the entry’s dirty bit. Clean entries are discarded immediately. Dirty entries are placed in the eviction buffer. Threads from the state thread pool pick an entry from the eviction buffer, write it to the backend, and remove it from the buffer.
The eviction buffer enables writes without blocking execution; when an entry is fetched from the backend, and the linked list is at full capacity, processing does not block waiting for a thread to get freed and perform the write of the evicted entry. If a read (or a \emph{prefetch hint}) concerns a key currently staged in the eviction buffer, the entry is moved back to the doubly-linked list.

\subsection{Interaction with Checkpointing and Watermark Techniques}
\label{sec:fault-tolerance-watermarks}
Keyed Prefetching is designed to be non-invasive to the underlying engine's fault-tolerance mechanisms, such as Apache Flink’s checkpointing mechanism~\cite{carbone2015lightweightasynchronoussnapshotsdistributed}. 
When a checkpoint is triggered, the engine ensures that all in-flight operators reach a consistent barrier; at this point, any modified state within the TAC---whether resident in the linked list or staged in the eviction buffer---is stored in the persistent state backend before the checkpoint is finalized. Because prefetch hints are treated as a side channel of the DAG, they are subject to the same alignment rules as regular records. Consequently, in the event of a failure, the consistent state is recovered from the last successful checkpoint in the persistent key-value store.

Further, Keyed Prefetching is inherently compatible with watermark-based mechanisms for handling out-of-order events~\cite{dataflow-model, spe-evolution-survey}. Watermarks track the progress of event time and define a threshold beyond which events are considered late. Since TAC orders entries by event timestamps, the state fetched for an out-of-order tuple is assigned the record's intrinsic event timestamp; this ensures the entry maintains its logical temporal priority in the cache, receiving the same treatment it would if it had arrived normally. Regarding prefetching hints, if a record is delayed before the source operator, both lookahead and stateful operators experience the same relative delay, and prefetching functions normally. If delayed internally between operators, prefetching remains successful, though the state will reside in the cache longer before being used.
Given that out-of-order events typically represent a small fraction of the total data volume, the overall impact on cache efficiency remains minimal. 
To optimize for out-of-order events, the \emph{Prefetching Manager} can be configured to discard hints for late tuples. Specifically, it could drop hints for records with timestamp older than the current watermark minus the lateness threshold, mirroring the behavior of the operators and avoiding unnecessary prefetching for records that will be dropped.

\section{Implementation}
\label{sec-implementation}

We implement Keyed Prefetching in the widely used Apache Flink~\cite{flink-original-paper} stream processing engine (v2.0). Following the design in Section~\ref{sec-method}: (i) the \emph{Hint Extractor} runs in \emph{lookahead} operators, (ii) the \emph{Prefetching Manager} is collocated with each stateful operator, and (iii) the \emph{Timestamp–Aware Cache} is placed in front of the state backend. We integrate the \emph{Hint Extractor} and the \emph{Prefetching Manager} inside Flink’s \emph{TaskManager}, while the \emph{Prefetching Controller} resides in the \emph{JobManager}.
Keyed Prefetching is compatible with any key-value store backend that supports concurrent accesses.

\textbf{Operators with multiple states}
Some operators maintain more than one keyed state (for example, a two-input join storing two hash tables, one for each input stream). For such operators, we use composite keys consisting of the key and the state identifier, which disambiguate which internal state an access targets.

\section{Evaluation}
\label{sec-experiments}

This section experimentally evaluates the effectiveness of Keyed Prefetching in reducing tail latency for stateful queries. We additionally report the achieved throughput and analyze its individual components.

\paragraph{Baselines} We compare Keyed Prefetching against three baselines: 
\begin{enumerate}
    \item \emph{Cache-LRU:} There is a Least-Recently-Used (LRU) cache in front of the key-value store.
    \item \emph{Cache-Clock:} There is a Clock cache in front of the key-value store.
    \item \emph{Asynchronous I/O:} State is fetched asynchronously, and while waiting on I/O, stateful operators move on with processing the next tuple.
\end{enumerate}
We implement all baselines in Apache Flink and use RocksDB as the state backend. The Asynchronous I/O baseline uses a cache as well---we report results for the best performing cache eviction policy between LRU and Clock.

\paragraph{Platform} We use 2 two-socket Intel Xeon Gold 5118 CPU (2.30 GHz) servers, each with ($12\times2$) threads per socket, $256$ GB of DRAM, and a 1.6 TB NVMe SSD (Dell Express Flash PM1725a AIC) for the TaskManagers. Additionally, we use a two-socket Intel Xeon E5-2680 v4 CPU (2.40 GHz) server with ($2\times14$) threads per socket for Flink's JobManager and a two-socket Intel Xeon E7-8890 v3 CPU (2.50GHz) server with ($2\times18$) threads per socket for data generation. Finally, we use an Intel Xeon E5-2650L v3 CPU (1.80GHz) with ($2\times12$) threads per socket for the Redis key-value store used for the Yahoo Streaming Benchmark.

\begin{figure}[t]
 \centering
    \includegraphics[width=0.9\linewidth]{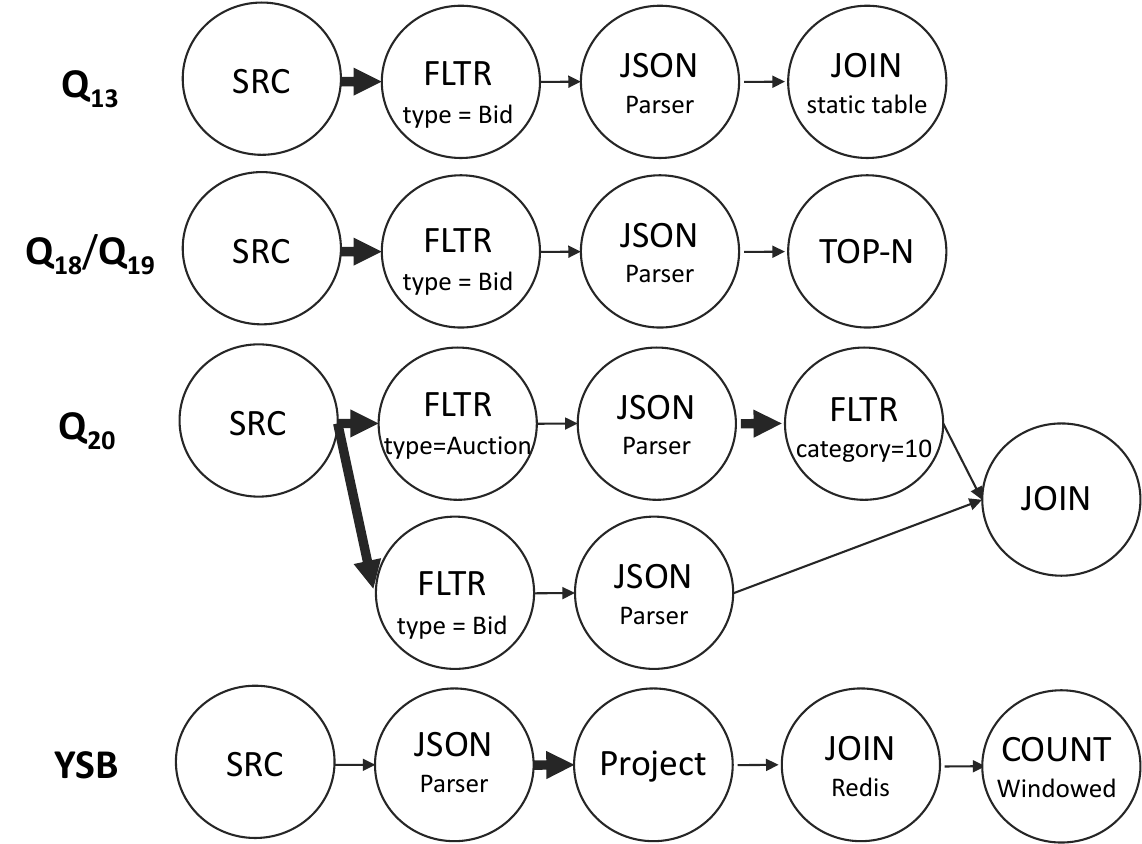}
    \caption{Dataflow graphs of the evaluated queries. Thicker arrows connect pipelined operators.}
	\label{fig:exp:dags}
\end{figure}

\begin{figure*}[t]
\centering
\begin{subfigure}[t]{0.32\linewidth}
\includegraphics[width=1\textwidth, keepaspectratio]{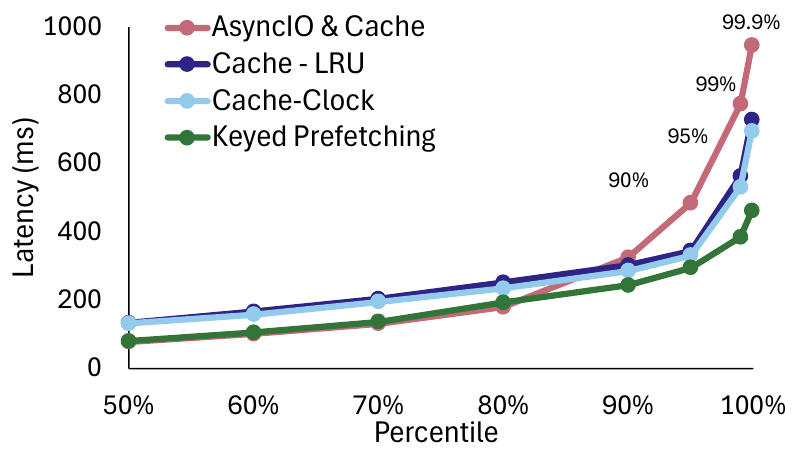}
\caption{Q13 (200 GB)}
\label{fig:exp:percentiles-q13} 
\end{subfigure}
\centering
\hfill
\begin{subfigure}[t]{0.32\linewidth}
\includegraphics[width=1\textwidth, keepaspectratio]{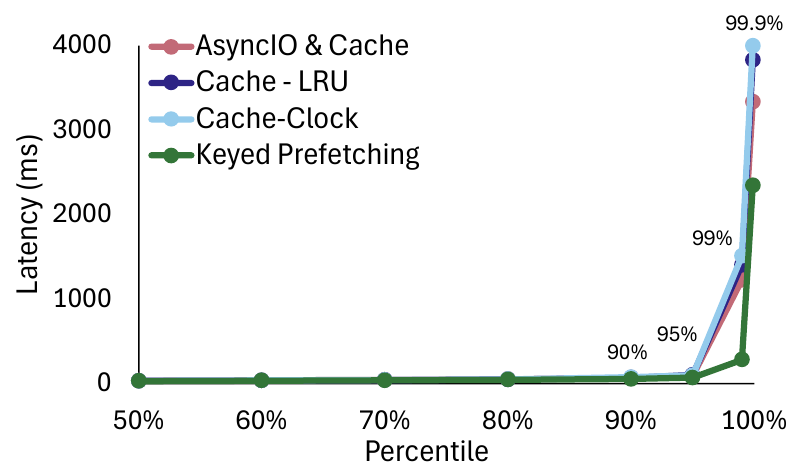}
\caption{Q18 (205 GB)}
\label{fig:exp:percentiles-q18}
\end{subfigure}
\centering
\hfill
\begin{subfigure}[t]{0.32\linewidth}
\includegraphics[width=1\textwidth, keepaspectratio]{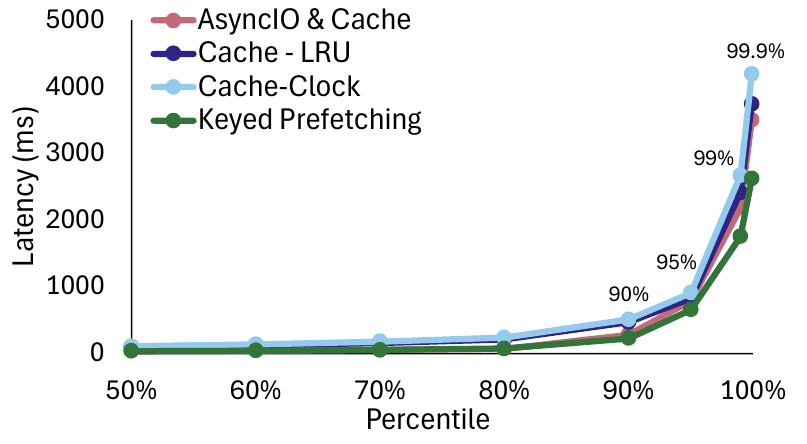}
\caption{Q19 (216 GB)}
\label{fig:exp:percentiles-q19}
\end{subfigure}
\centering
\begin{subfigure}[t]{0.32\linewidth}
\includegraphics[width=1\textwidth, keepaspectratio]{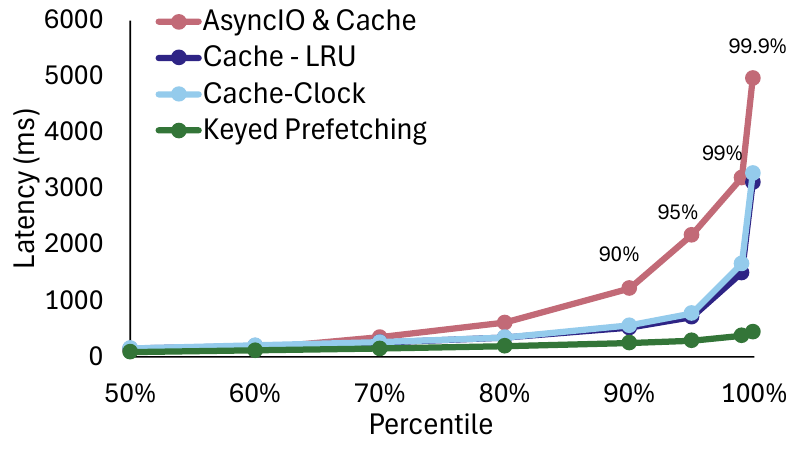}
\caption{Q20 (197 GB)}
\label{fig:exp:percentiles-q20}
\end{subfigure}
\centering
\begin{subfigure}[t]{0.32\linewidth}
\includegraphics[width=1\textwidth, keepaspectratio]{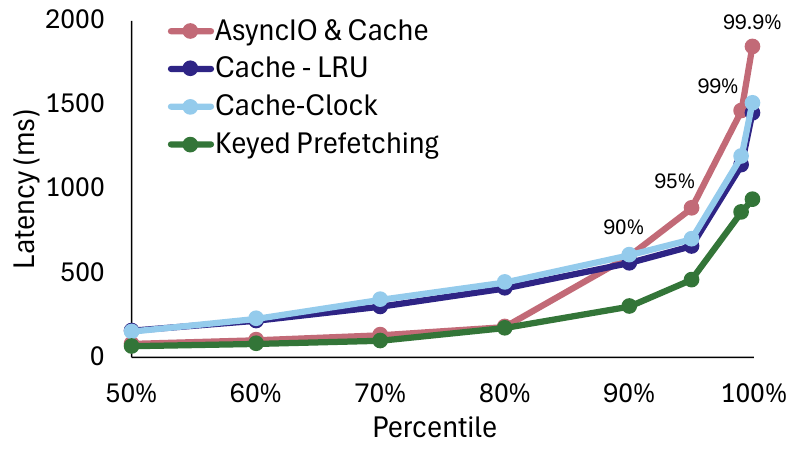}
\caption{YSB (220 GB)}
\label{fig:exp:percentiles-ysb}
\end{subfigure}
\caption{\label{fig:exp:percentiles} End-to-end percentile latency. In the captions, for each query, we provide an estimate of the state size.}
\end{figure*}

\paragraph{Methodology} 
We adopt the methodology of prior work on benchmarking streaming systems~\cite{DBLP:conf/icde/KarimovRKSHM18, agnihotri2025pdspbenchbenchmarkingparalleldistributed}, ensuring that data generation runs on a separate machine to eliminate interference between data production and processing. The data generator is producing events in JSON format. For NEXMark, we use a single generator for all event types~\cite{nexmark-beam} and the event type and primary/foreign keys are represented as top-level attributes. We allow the system to warm up and collect measurements only after reaching a steady state and after the state exceeds the cache size.

\paragraph{Workloads}
For our experiments, we use two benchmarks:
(i) The \textbf{NEXMark benchmark}~\cite{nexmark-beam, Tucker2002NEXMarkA} emulates online auctions and is used to benchmark SPEs~\cite{flowKV2023, apache-beam-ibm, kalavri-three-steps, Flink-v2-release}. It has three types of input tuples: Auction, Bid, and Person. We select the queries that generate large states~\cite{Flink-v2-release}. Concretely, we evaluate four queries: Q13 joins the Bid stream to a bounded side input based on auctionId, modeling tuple enrichment. Q18 finds the top-1 bid for an auction based on the bidding time. Q19 outputs the current top-10 bids for an auction based on price. Q20 performs a join between the Auction and Bid streams and uses a filter on auctions ($category=10$). Q13, Q19, Q20 organize state based on auction ID, and Q18 based on (auction ID, bidder ID). Q7 also generates a large state, but we exclude it because the stateful operator accesses state only via append operations for input tuples and read operations at the end of the window---as explained in Section~\ref{sec-overview}, these access patterns are not targeted by Keyed Prefetching. The average byte-serialized size of Person, Auction, and Bid tuples is 200B, 500B, and 200B, respectively. Data streams created by the generator consist of 2\% person tuples, 6\% auction tuples, and 92\% bid tuples. Regarding NEXMark's data generation parameters, we configure auctions and bidders to remain active for a 2-hour window, and the most popular auction/bidder changes every second to allow for large active states. We use the default skewness parameters specified in the open-source implementation~\cite{nexmark-beam}; a Bid corresponds to a hot auction with probability 50\% (this parameter affects the distribution of the state-access key of all queries) and to a hot bidder with probability 75\% (this parameter affects the distribution of the state-access key of Q18). We vary skewness in Section~\ref{sec:exp:skew}.
(ii) The \textbf{Yahoo Streaming Benchmark (YSB)}~\cite{ysb} provides a scenario where the engine must access disaggregated state over the network. 
It emulates an ad analytics pipeline: a stream of ad events is enriched through a join with campaign metadata stored in a remote Redis key-value store. The state access key is the ad ID. We use a Zipfian distribution ($\alpha= 1$) to generate ad IDs. The byte-serialized size of each event is 114B.

Fig.~\ref{fig:exp:dags} presents the dataflow graphs of all queries. For Nexmark, we pipeline the event-type filter with the source to avoid excessive communication. The JSON parser operator is not pipelined with the source to fully exploit available parallelism; the parallelism of the source is capped at 16 (the maximum number of client connections supported by the data generator). Q18 and Q19 have the same query plan structure but differ in the configuration of the TopN operator.

\paragraph{Configuration}
We run a single Flink TaskManager with 24 task slots on each worker node. 
We use a total of 128GB of memory (per task manager, 50GB is allocated for caching operator state, 5GB is dedicated to RocksDB for memtables, and indexes, and the rest is used by Flink for network buffers and framework memory). 
We follow the official tuning guideline for RocksDB~\cite{rocksdb-tuning}. 
To avoid interference from the OS page cache, we enable Direct I/O for RocksDB.

Following guidelines on configuring Apache Flink for low tail latency~\cite{low-latency-config-flink}, we set the buffer timeout interval to 30 milliseconds and the input rate to the maximum sustainable while keeping the average CPU utilization below 70\%. For Keyed Prefetching, we configure the Count-Min sketch frequency filter with $d=4$, $w=10000$, $b=8$, and $T=20$, using an aging interval of $\Delta=1000$ records. Fig.~\ref{fig:exp:freq-filter} illustrates the impact of varying the threshold $T$ on end-to-end latency. We inject a marker for the adaptive lookahead selection mechanism every 100 ms.

\subsection{End-to-end percentile latency}
\label{sec:exp:percentiles}

Fig.~\ref{fig:exp:percentiles} presents the end-to-end latency for each approach. 
For each query, we provide in the caption an estimate of the state size using access pattern knowledge and the total number of tuples processed. These estimates concern only operator state for NEXMark and the state size of the Redis store for YSB.  

Cache-LRU and Cache-Clock perform similarly across all queries as they result in similar cache-hit ratios. As expected, Asynchronous I/O results in lower p50 latency than caching. Concretely, it reduces p50 latency by $1.02-2.63\times$ across the different queries. However, for higher percentiles, the latency of Asynchronous I/O increases, and for Q13, Q20, and YSB, it exceeds the latency achieved by caching methods. This happens because Asynchronous I/O adds overheads as it needs to handle many concurrent threads accessing state and check for completed fetches. These overheads become more prominent and can outweigh the benefits of Asynchronous I/O for high percentiles. For Q18 and Q19, where the amount of state accesses per operation (reads and writes) is higher, Asynchronous I/O pays off even in high percentiles.

Q13 exhibits only read operations and experiences lower latency in high percentiles than the other queries. Q18 has overall more keys that are frequent at any point in time. Additionally, it has a higher total number of keys, which results in many tuples corresponding to the first occurrence of their key; these tuples are served quickly thanks to RocksDB's bloom filters. Both these factors result in latency that increases more slowly up to the 95th percentile. 

Both Q13 and YSB perform tuple enrichment by joining with static data. Although they are not directly comparable due to differences in the rest of their query plans, data distribution, and event sizes, a primary factor contributing to YSB's higher overall latency is its disaggregated state architecture.

Keyed Prefetching reduces latency at high percentiles for all queries while achieving p50 latency similar to or lower than that of Asynchronous I/O. Concretely, Keyed Prefetching lowers p50 latency by $1.21\times$, $1.07\times$, and $1.2\times$ for Q20, Q19, and YSB, respectively, and has less than 3 milliseconds higher p50 latency for Q18 and Q13 compared to Asynchronous I/O. For Q20, Keyed Prefetching results in a higher speedup than for other queries; Q20 processes two input streams (Auction and Bid). Most events are bids and, with Keyed Prefetching, are almost exclusively served from the cache as they access the relatively small state formed by the Auction stream, which mostly resides in cache thanks to the prefetching hints. Some events need to access cold state, but there is enough time for Keyed Prefetching to hide these accesses.

\textbf{Takeaway:} Keyed Prefetching can hide expensive state accesses to local and disaggregated backends and reduces p999 latency by $1.34-11\times$ compared to the baselines while resulting in the same or lower p50 latency.

\subsection{Varying degrees of skewness and cache size}
\label{sec:exp:skew}

\begin{figure}[t]
\centering
\begin{subfigure}[t]{0.9\linewidth}
\includegraphics[width=1\textwidth, keepaspectratio]{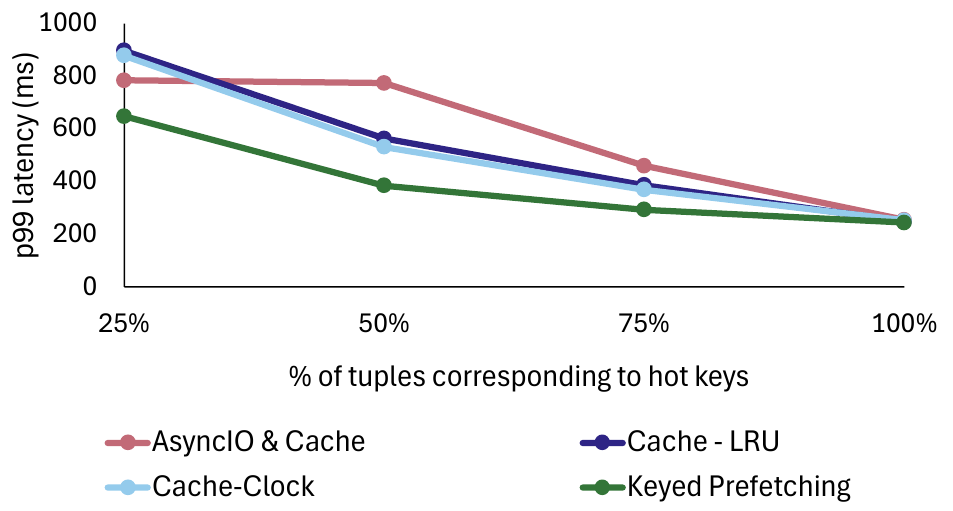}
\caption{p99 latency}
\label{fig:exp:skewp99} 
\end{subfigure}
\hfill
\begin{subfigure}[t]{1\linewidth}
\includegraphics[width=0.9\textwidth, keepaspectratio]{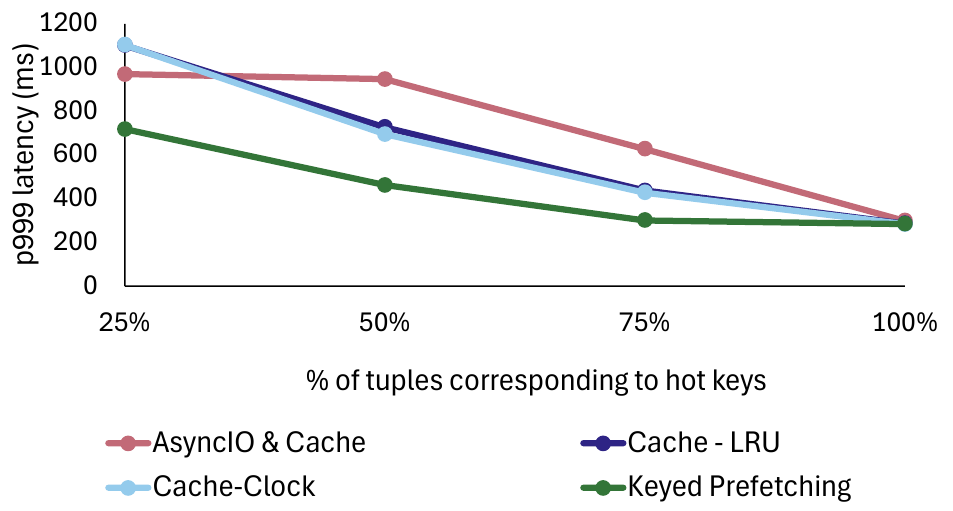}
\caption{p999 latency}
\label{fig:exp:skewp999}
\end{subfigure}
\caption{\label{fig:exp:skew} p99 and p999 latency of Q13 with varying percentages of tuples that correspond to hot keys}
\end{figure}

Fig.~\ref{fig:exp:skew} shows the p99 and p999 latency of Q13 as the percentage of bid events that correspond to hot auctions changes from $25\%$ to $100\%$.
Tail latency drops as skewness increases because the benefit of decreasing cache misses outweighs the effects of load imbalance.
At a low $25\%$ percentage, Asynchronous I/O results in lower tail latency than caching methods. However, as skewness increases, Asynchronous I/O results in higher tail latency compared to caching. When all tuples correspond to hot auction IDs, all approaches perform similarly since the entire active state fits in memory.
Keyed Prefetching outperforms all baselines across the different skewness degrees.

\begin{figure}[t]
\centering
\begin{subfigure}[t]{1\linewidth}
\includegraphics[width=0.9\textwidth, keepaspectratio]{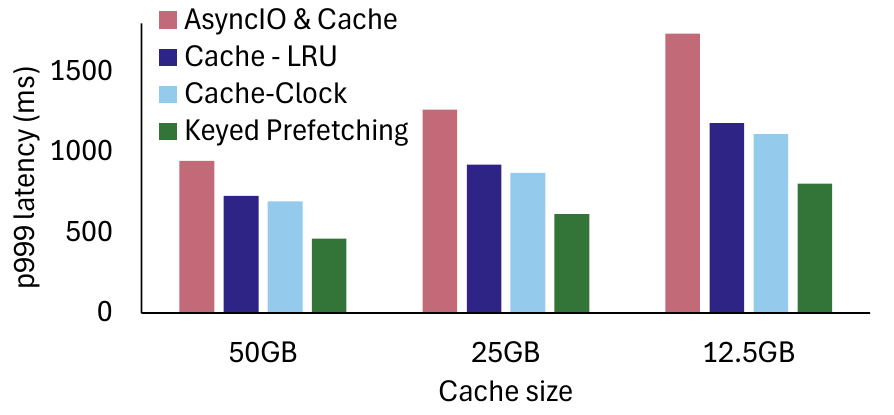}
\caption{Q13}
\label{fig:exp:cache-size-q13} 
\end{subfigure}
\hfill
\begin{subfigure}[t]{1\linewidth}
\includegraphics[width=0.9\textwidth, keepaspectratio]{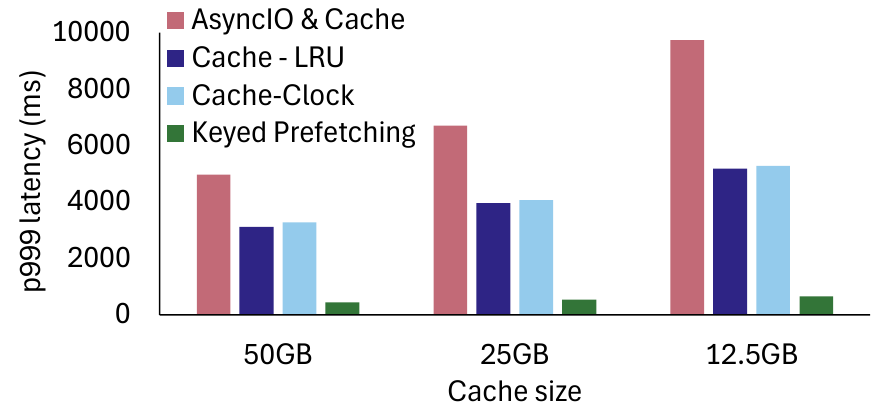}
\caption{Q20}
\label{fig:exp:cache-size-q20}
\end{subfigure}
\caption{\label{fig:exp:cache-size} p999 latency with varying cache sizes}
\end{figure}

Fig.~\ref{fig:exp:cache-size} shows the p999 latency as we vary the cache size per task manager for Q13 and Q20 (the other workloads lead to the same insights). Asynchronous I/O results in a higher latency increase than caching methods because, as the cache size decreases, there is greater pressure on the threads used to fetch state. Keyed Prefetching outperforms all baselines across the different configurations. For Q20, Keyed Prefetching is affected less by the cache size than for Q13 because a small part of the state is very active, and thanks to the \emph{prefetch hints} and the efficiency of the \emph{Timestamp-Aware Cache}, it stays in the cache even with smaller cache sizes.

\textbf{Takeaway:} Keyed Prefetching consistently outperforms all baselines for varying degrees of skewness and cache sizes.

\subsection{Robustness and overhead of Keyed Prefetching}
\label{sec:exp:overhead}

\begin{figure}[t]
 \centering
    \includegraphics[width=0.9\linewidth]{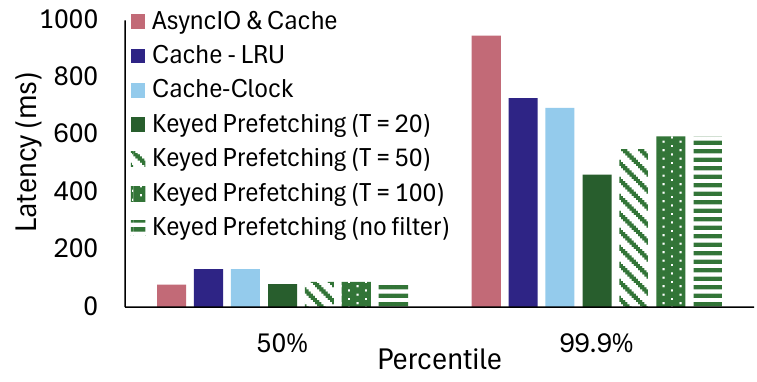}
    \caption{Impact of CMS threshold $T$ on latency (Q13). Increasing $T$ raises the frequency bar for hot-key classification, increasing the number of hints. ``no filter'' corresponds to sending prefetching hints for all tuples.}
	\label{fig:exp:freq-filter}
\end{figure}

In Fig.~\ref{fig:exp:freq-filter}, we show the results from evaluating the impact of the CMS frequency filter threshold ($T$) on the end-to-end latency of Q13.
Higher values of $T$ translate to a stricter threshold for a key to be characterized as hot, thereby increasing the volume of \emph{prefetch hints}.  Our results show that while the p50 latency remains relatively unaffected, tuning the threshold has visible impact on p999 latency. Importantly, all evaluated configurations outperform both Asynchronous I/O and caching methods at high percentiles, while maintaining a p50 comparable to Asynchronous I/O. A promising direction for future work is to dynamically tune the threshold at runtime based on the observed cache miss ratio.

\begin{figure}[t]
 \centering
    \includegraphics[width=0.8\linewidth]{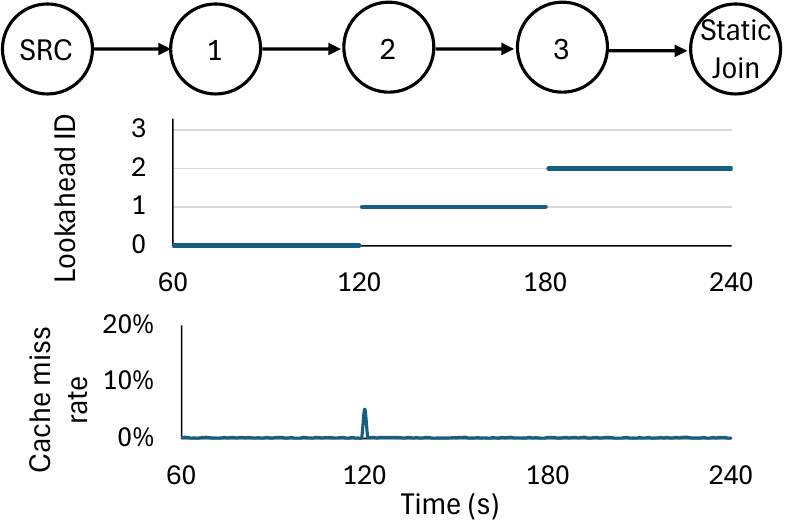}
    \caption{Dynamic lookahead adaptation under varying hint accuracy and state access latency. At time 120s, operator 1 starts changing the distribution of the state access key. At time 180s, the state access latency of the static join decreases.}
	\label{fig:exp:lookahead}
\end{figure}

Fig.~\ref{fig:exp:lookahead}, showcases the responsiveness of our dynamic lookahead selection mechanism to fluctuations in hint accuracy and latency. Specifically, we use a synthetic query emulating a stateful join with a static table, which allows us to control the state access latency. We insert three UDF operators upstream of the join, all serving as candidate lookaheads. We allow the system to warm up for 1 minute. Keyed Prefetching initially selects the source operator as the lookahead ($ID = 0$). At 120s, operator 1 starts changing the distribution of the state access key. 
The \emph{Prefetching Manager} detects the drop in hint accuracy and shifts to lookahead 1. This transition causes a momentary spike in the cache miss ratio because no hints are sent for a few tuples during the switch. At 180s, the state access latency of the static join decreases, and the prefetching manager switches to lookahead 2, which now provides more timely hints. No increase in cache misses occurs here; some hints are duplicated during the shift, but no hints are missed. Overall, Keyed Prefetching quickly adjusts the lookahead operator to maintain a low cache miss ratio thanks to its lightweight marker-based coordination. The increase in cache misses occurs for operators that change the state access key distribution in a way that introduces new keys. An operator that only drops keys (e.g., a filter) would not increase cache misses but only lead to prefetching misses.

\begin{table}[htbp]
\caption{Impact on CPU}
\label{table:reconfiguration}
\begin{center}
\begin{tabular}{|c|c|c|c|c|c|}
\hline
\textbf{} & \textbf{\textit{Q13}} & \textbf{\textit{Q18}} & \textbf{\textit{Q19}} & \textbf{\textit{Q20}} & \textbf{\textit{YSB}} \\
\hline
 \textbf{Cache-LRU}  & 61.5\% & 68.4\% & 69.8\% & 61.7\% & 63.6\%\\ 
 \hline
 \textbf{Cache-Clock} & 60.7\% & 68.9\% & 70.7\% & 62.5\% & 64.3\%\\
 \hline
 \textbf{AsyncIO \& Cache}  & 44.1\% & 42.5\% & 46.9\% & 30.4\% & 35.5\%\\ 
 \hline
\textbf{Keyed Prefetching} & 43.5\% & 42.2\% & 46.3\% & 29.3\% & 35.0\%\\
 \hline
\end{tabular}
\label{tab-cpu}
\end{center}
\end{table}

\begin{table}[htbp]
\caption{Network overhead of Keyed Prefetching over Cache-LRU}
\label{table:reconfiguration}
\begin{center}
\begin{tabular}{|c|c|c|c|c|c|}
\hline
\textbf{} & \textbf{\textit{Q13}} & \textbf{\textit{Q18}} & \textbf{\textit{Q19}} & \textbf{\textit{Q20}} & \textbf{\textit{YSB}} \\
\hline
 \textbf{Keyed Prefetching} & 2.2\% & 1.8\% & 2.4\% & 2.7\% & 8.7\%\\
 \hline
\end{tabular}
\label{tab-net}
\end{center}
\end{table}

Table~\ref{tab-cpu} reports the CPU utilization using Apache Flink's \emph{busyTimeMsPerSecond} metric. This metric accounts for the total time a thread is occupied, which includes cycles spent waiting for I/O or network communication. As mentioned in Section~\ref{sec-experiments}(e), we set the input rate to the maximum sustainable level that maintains utilization below 70\%; consequently, the caching baselines operate near this limit. Asynchronous I/O and Keyed Prefetching exhibit significantly lower utilization at the same input rate as they overlap computation with state accesses.
Table~\ref{tab-net} reports the network overhead of Keyed Prefetching. While reported against Cache-LRU, these results generalize to all baselines as they share similar network profiles. Q18 exhibits the lowest overhead due to higher key skewness, translating to fewer prefetch hints. YSB has higher overhead as its data tuples are smaller and the prefetch hints have larger payloads. Finally, regarding memory usage, all approaches share the same memory budget. Keyed Prefetching dedicates a portion of this budget to the lightweight CMS frequency filters (40KB per lookahead subtask). Given the number of lookahead operators and their parallelism, this results in a total filter footprint of 640KB for Q13, Q18, and Q19, 1600KB for Q20, and 960KB for YSB.

\textbf{Takeaway:} Keyed Prefetching consistently outperforms the baselines regardless of filter tuning, exhibits low overhead, and dynamically adapts the lookahead operator at runtime. 

\subsection{Impact of Keyed Prefetching on Throughput}
\label{sec:exp:tput}

\begin{figure}[t]
 \centering
    \includegraphics[width=1\linewidth]{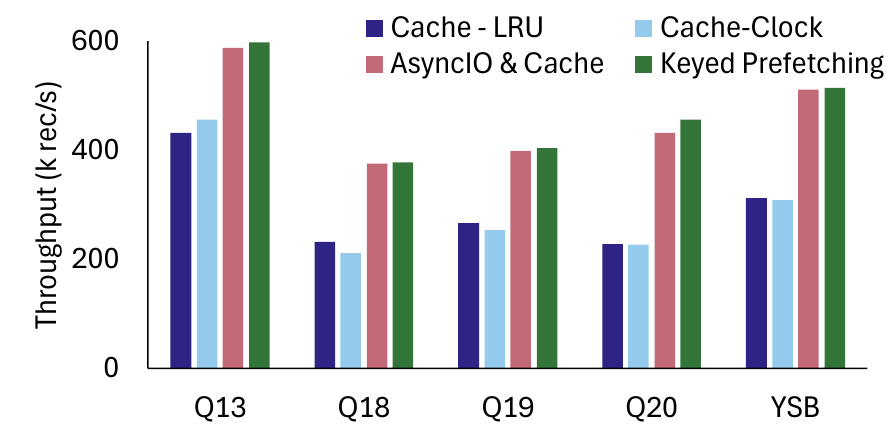}
    \caption{Maximum sustainable throughput}
	\label{fig:exp:tput}
\end{figure}

Fig.~\ref{fig:exp:tput} shows the impact of the different approaches on query throughput for all queries. We report the maximum sustainable throughput the various techniques achieve without accumulating backpressure. Cache-LRU and Cache-Clock have similar throughput across queries. Asynchronous I/O improves throughput by overlapping state access with computation. Keyed Prefetching has $1.01-1.2 \times$ higher throughput than Asynchronous I/O. 

\textbf{Takeaway: } Keyed Prefetching not only minimizes latency but also can increase throughput by $1.01-2\times$ compared to the baselines.

\section{Related Work}
\label{sec:related}

\textbf{Large State Management for Streams}
Recent works have studied how to manage large-scale state of streaming applications and have pointed out possible performance problems~\cite{benchmark-stream-stores, rhino, kalavri-2020}. These works motivated Keyed Prefetching, a novel query-plan-aware prefetching technique. Gadget~\cite{benchmark-stream-stores} proposes a benchmark for evaluating state stores in stream processing. Rhino~\cite{rhino} is a novel technique for efficient state migration and fault recovery for states that exceed main memory. Meces~\cite{meces} lowers latency during rescaling of stateful streaming jobs by prioritizing the migration of hot keys.

\,

\textbf{Prefetching}
Existing prefetching techniques rely on predictable patterns, perform speculative fetches, and traditionally operate within the scope of a single processor's memory accesses. 
Software prefetching techniques allow programmers or compilers to insert prefetch instructions~\cite{prefetching-joins, btreeprefetch, fractalPrefetch, 10.1145/237090.237190, 10.1145/143371.143488}. 
Previous work has addressed database-specific prefetching (e.g., for joins~\cite{prefetching-joins} and B+trees\cite{btreeprefetch, fractalPrefetch}).
Keyed prefetching is specifically designed for streaming applications. It operates at the granularity of keys, utilizes timing information inherently available in streaming systems, and targets low tail latency. Additionally, it operates in a distributed environment by passing information across operators in the query's dataflow graph.

FlowKV~\cite{flowKV2023} proposes a persistent store tailored for streaming. It features a prefetching optimization for fetching window state of queries using unaligned windows (i.e., windows triggering at different points in time) by predicting when a window will fire.
This optimization is complementary to our prefetching technique, which targets state accesses that happen during the window upon a tuple's arrival.

Palpatine~\cite{palpatine} is an in-memory application-level cache for distributed key-value data stores that employs data-mining techniques to predict future access patterns and prefetch data.
Palpatine targets general storage workloads and employs probabilistic prefetching. Therefore, when a lookahead operator is available to provide exact hints, Keyed Prefetching is the preferred choice. In the absence of such an operator, speculative prefetching approaches like Palpatine could be employed, provided their mining overhead does not compromise real-time streaming constraints.

EIRES~\cite{eires-zhao} combines cost-guided prefetching and lazy evaluation to integrate remote static data into Complex Event Processing (CEP) workflows. As it targets scenarios where access is probabilistic, since most partial matches are discarded, it employs a cost model to decide whether to prefetch. Keyed Prefetching hides latency in cases where a lookahead operator providing exact hints exists, whereas EIRES is better suited for managing uncertain access patterns in event pattern matching.

\,

\textbf{Information passing} 
Information passing—pushing knowledge learned in one part of a plan to guide work elsewhere—has been used to avoid wasted I/O and computation. Concretely, existing work has studied passing join-derived information from one operator/table to another to prefilter data earlier in the plan~\cite{parachute, looking-ahead-zhu, predicate-caching, sideways-inf-passing}. For example, Sideways-information passing~\cite{sideways-inf-passing} builds bloom filters on the build sides and pushes them into the upcoming pipelines to deal with the issue of expensive hash-table probes.
Keyed Prefetching employs information passing to hide streaming state I/O.

\,

\textbf{Low-latency stream processing}
The problem of reducing latency in stream processing has been extensively studied~\cite{8514883, Tail-latency-du, tail-latency-socc, shadowsync, replication-latency, 10.1007/s10586-022-03758-1}. For example, previous work aims to reduce latency caused by load imbalance\cite{Tail-latency-du, tail-latency-socc}, and node failures and slowdowns~\cite{replication-latency}.
ShadowSync~\cite{shadowsync} addresses the problem of increased latency caused by maintenance LSM-tree operations.
Our work addresses the problem of increased tail latency caused by accessing large states.

\,

\textbf{Caching}
Previous work~\cite{patterson-1995, palpatine} uses a separate cache for prefetching and previously used elements, and based on cost analysis, balances the two. In contrast, our \emph{Timestamp-Aware Cache}, designed specifically for streaming, unifies the two different state types by ordering both based on timestamps.
Samza~\cite{samza-linkedin}, a distributed system for stateful and fault-tolerant stream processing, reduces the latency of accessing the state of persistent key-value stores by placing a cache in front of the key-value store. This is similar to the caching baselines we compared Keyed Prefetching to (Section~\ref{sec-experiments}).

\section{Conclusion}

Stream-processing applications often execute time‑critical tasks under strict latency SLOs.
Any non-trivial streaming computation includes stateful operators. As data volume and query complexity grow, and applications transition to cloud environments, streaming systems must handle state that lives outside memory, on local storage, or in disaggregated backends. Streaming systems traditionally couple state access with the data path, placing expensive state access in the critical path of execution and inflating tail latency. 
We propose Keyed Prefetching, which decouples state access from the data path by extracting future access keys at upstream \emph{lookahead} operators and fetching the corresponding state as soon as the key is known, rather than when the tuple reaches the stateful operator.
Keyed Prefetching is accurate---\emph{lookahead} operators send exact hints---and timely---it selects the \emph{lookahead} whose hints arrive early enough to hide I/O. To use memory efficiently, we introduce Timestamp‑Aware Caching, which jointly manages previously accessed and prefetched entries using event timestamps as a single ordering signal. In our evaluation, Keyed Prefetching reduces p999 latency by $1.34-11\times$, while maintaining or increasing throughput.

\section{AI-Generated Content Acknowledgement}
OpenAI’s ChatGPT was used for proofreading. We asked the system to flag potential grammatical errors, typos, and non-idiomatic phrasing in the text and to propose fixes.

\bibliographystyle{IEEEtran}
\bibliography{bib}

\clearpage

\end{document}